\documentstyle[aps,amssymb,multicol]{revtex}
\def\be{\begin{equation}}
\def\ee{\end{equation}}
\def\bq{\begin{eqnarray}}
\def\eq{\end{eqnarray}}
\def\Rc{\check{R}}
 1

\begin{document}
\draft
\title{Integrable Extended Hubbard Hamiltonians from Symmetric Group
Solutions}

\author{Fabrizio Dolcini
and Arianna Montorsi}
\address{Dipartimento di Fisica and Unit\`a INFM, Politecnico di Torino,
I-10129 Torino, Italy}
\date{22 November 1999}
\maketitle
\begin{abstract}
We consider the most general form of extended Hubbard Hamiltonian
conserving the total spin and number of electrons, and find all the
1-dimensional completely integrable models which can be derived from
first degree polynomial solution of the Yang-Baxter equation. It is
shown that such models are 96. They are identified with the
16-dimensional representation of a new class of solutions of symmetric
group relations, acting as generalized permutators. As particular
examples, the EKS and some other known models are obtained. 
\pacs{1998 PACS number(s): 71.10.Pm; 71.27.+a; 05.30.-d }
\end{abstract}

\begin{multicols}{2}
Recently the interest in low-dimensional correlated electron systems has 
motivated some attention on the search for completely integrable
1-dimensional extended Hubbard models \cite{EKS} $\div$ \cite{SCSH}.
The physical relevance of the latter has been realized in very
different context, from high-$T_c$ materials\cite{HEN} to quantum
wires\cite{LIGR}. Determining the integrability of model Hamiltonians
can yield a wide set of rigorous results, very rare for interacting
many-body systems, as the exact ground state phase diagram and the
evaluation of correlation functions. In general, the complete
integrability of a model is proved by providing the underlying
$R$-matrix, solution of the Yang-Baxter equation (YBE), which is
related to the factorizability of the scattering matrix; whereas its
spectrum can be obtained by diagonalizing the corresponding Transfer
matrix. To achieve both the above purposes, the Quantum Inverse
Scattering Method (QISM, see \cite{KBI} and references therein) has
turned out to be a very powerful tool. Within this scheme, the standard
approach is to impose some appropriate (quantum)symmetries onto the
$R-$matrix, and then to derive from it the integrable Hamiltonian
sharing the same symmetries. Whether the model so obtained does
actually correspond to a physical system or not, it is always an
{\it a posteriori} check to do.
\\On the contrary, when having a specific physically interesting
Hamiltonian $H$ as input, one may wonder whether a $R$-matrix can 
be obtained as an output. To this
purpose, in a recent paper \cite{DOMO} it was noticed how one can
explicitly build up the $R-$matrix as a polynomial of degree $p$ in the
spectral parameter, with the requirement that its first derivative is
precisely $H$. Constraints on the coefficients of the power expansion
were obtained, by imposing the YBE, in the standard form depending on
the difference of spectral parameters. In particular, it was shown that,
in all cases in which the expansion was finite, the highest $3 p$-order
equation is the YBE with no spectral parameter, {\sl i.e.}
the braid-group relation \cite{HIET}, whereas the $2 p$-order
specializes this relation to the symmetric group case.

Starting from this observation, in the present letter we consider the
most general form of the (isotropic) extended Hubbard model
preserving the spin components and the number of electrons. By imposing
that the local Hamiltonian satisfies the symmetric group relations we
derive the constraints among the physical parameters which guarantee its
integrability, the $R$-matrix simply being in this case a first-degree
polynomial. In so doing we both obtain some known results, for instance
the $EKS$ model Hamiltonian, and prove the integrability of many other
cases. It is realized that the solutions act as generalized 
permutators (see below) on the chain sites. At the same time, we find
all the solutions of symmetric group relations having the form of a
generalized permutator, and show that the above integrable extended
Hubbard models are precisely those symmetric group generalized
permutators (in the 16-dimensional representation) that preserve the 
Hamiltonian symmetries. 
\\Interestingly our results are obtained without imposing any
extra symmetry constraints. Nevertheless, like in the quantum symmetric
cases, they have a quite simple structure, deriving from the underlying
structure of generalized permutator solutions of the symmetric group 
relations.

The most general form of the extended Hubbard model \cite{HUB},
invariant under spin-flip and conserving the total number of electrons
and the magnetization, was considered in some papers
\cite{dBKS}. Its Hamiltonian reads
\end{multicols}
\begin{eqnarray}
&{\cal H}& = \mu_e\sum_{{j},\sigma}
n_{{j},\sigma} - \sum_{<{j},{k}>,\sigma }[t-X (n_{{j},
-\sigma}
+ n_{{k},-\sigma})+\tilde X n_{{j},-\sigma}
n_{{k},-\sigma} ] c_{{j},\sigma}^\dagger
c_{{k}, \sigma} +  U \sum_{j} n_{{j},\uparrow}n_{{j},\downarrow} +
{V\over 2}\sum_{<{j},{k}>} n_{j} n_{k} \label{ham}  \\
&+& {W\over 2} \sum_{<{j},{k}>,\sigma,\sigma' }
c_{{j},\sigma}^\dagger c_{{k},
\sigma'}^\dagger c_{{j},{\sigma}'} c_{{k}, \sigma}
+ Y\sum_{<{j},{k}> } c_{{j},\uparrow}^\dagger
c_{{j},\downarrow}^\dagger  c_{{k},\downarrow} c_{{k},
\uparrow}
+ P \sum_{<{j},{k}> } n_{{j},\uparrow} n_{{j},
\downarrow} n_{k} + {Q\over 2} \sum_{<{j},{k}> }
n_{{j},\uparrow} n_{{j}, \downarrow} n_{{k},\uparrow}
n_{{k}, \downarrow} 
\, , \nonumber
\end{eqnarray}
\begin{multicols}{2}
\noindent where $\mu_e$ is the chemical potential,
$c_{{j},\sigma}^\dagger , c_{{j},\sigma} \,$ are fermionic
creation and annihilation operators ($ \{ c_{{j},{\sigma}'},
c_{{k},\sigma}\} = 0 \,$, $\, \{ c_{{j},\sigma} , c_{{k},
\sigma'}^\dagger \} = \delta_{{j},{k}}\, \delta_{\sigma,\,
\sigma'} {\Bbb I}$, $\, n_{{j},\sigma} \doteq c_{{j},
\sigma}^\dagger c_{{j},\sigma}$, $n_{j} = \sum_\sigma
n_{{j},\sigma}$ ) on a 1-dimensional chain $\Lambda$ (${j} ,
\, {k} \in \Lambda$, $\sigma \in \{ \uparrow , \downarrow \}$), and
$< {j} , \, {k} >$ stands for nearest neighbors ({\it n.n.}) in
$\Lambda$. In (\ref{ham})  the term $t$ represents the band
energy of the electrons, while the subsequent terms describe their
Coulomb interaction energy in a narrow band        approximation: $U$
parametrizes the on-site diagonal  interaction, $V$ the neighboring
site charge interaction, $X$ the  bond-charge interaction, $W$ the
exchange term, and $Y$ the pair-hopping  term. Moreover, 
additional many-body coupling terms have been included  in
agreement with \cite{dBKS}: $\tilde X$ correlates hopping
with on-site occupation number, and $P$ and $Q$ describe three- and
four-electron interactions. 
\\It is easily verified that $\cal{H}$ commutes with the global symmetry
algebra $su(2)\oplus u(1)$, where $su(2)$ is generated by
$S_+=\sum_{j} c_{{j},\uparrow}^\dagger c_{{j},\downarrow}$,
$S_-=S_+^\dagger$, $S_z={1\over 2}\sum_{j} (n_{{j},\uparrow}-
n_{{j}, \downarrow})$, and $u(1)$ corresponds to the conservation
of the total number of electrons ${\cal N}_e=\sum_{j} (n_{{j},
\uparrow} +  n_{{j},\downarrow})$.  

The Hamiltonian (\ref{ham}) acts on a vector space $V$ which is the
tensor product of 4-dimensional on-site spaces.
The local vector space $V_j$ associated with the $j-$th site is spanned
by the basis vectors $|\alpha\rangle_j = h^{(\alpha)}_j|o\rangle_j \;,
\; \alpha=1,.., 4 $ where $h^{(1)}_j=c^{\dagger}_{j\,\uparrow}  \, ; \, 
h^{(2)}_j=c^{\dagger}_{j\,\downarrow} \, ; \, h^{(3)}_j=1 \, 
; \, h^{(4)}_j = c^{\dagger}_{j\,\downarrow} 
c^{\dagger}_{j\,\uparrow} \, $ and $|o\rangle_j$ is the local 
vacuum ($c^{}_{j\sigma}|o\rangle=0 \;\; \forall \sigma$). Due 
to the fermionic nature of the creation operators, such vectors 
are endowed with an intrinsic grading $\epsilon$, {\it i.e.} 
$\epsilon(1)=\epsilon(2)=1$ and $\epsilon(3)=\epsilon(4)=0$.
The basis vectors of $V$ are defined as $| \alpha_1, 
\alpha_2 ,\ldots \alpha_N \rangle \, = \, h^{(\alpha_1)}_1 \ldots
h^{(\alpha_N)}_N | 0 \rangle$, with $|0\rangle$ 
the global vacuum ($c^{}_{j\sigma}|0\rangle=0 \; \forall \, j,\sigma$).
In order to work out calculations for fermionic Hamiltonians, it 
is useful to associate the basis vectors $e_{\alpha}$ of an ordinary 
${\Bbb C}^4$ euclidean space with the vectors $|\alpha\rangle_j$ 
of $V_j$, so that 
$|\alpha\rangle~\leftrightarrow~e_{\alpha}~=~(0,..,1,..,0)^T$, the 
1 being at the $\alpha$-th position. The $e_{\alpha}$'s inherit the 
grading $\epsilon$ of the vectors $|\alpha\rangle_j$.
Moreover, we shall also introduce a matrix representation $H$ for the 
Hamiltonian $\mathcal{H}$, whose elements are defined through 
${\mathcal{H}}\,| \beta_1, \beta_2 ,\ldots \beta_N \rangle=
H^{\alpha_1,\ldots, \alpha_N}_{\beta_1 \ldots ,\beta_N} \,| \alpha_1, 
\alpha_2 ,\ldots \alpha_N \rangle$.

${\cal H}$ is then said to be integrable if it exists a matrix $\Rc$,
acting on ${\Bbb C}^4\otimes {\Bbb C}^4$, which i) obeys the YBE,  
\be
\Rc_{23} (u-v) \Rc_{12} (u) \Rc_{23} (v)= \Rc_{12} (v) \Rc_{23}(u)
\Rc_{12} (u-v)
\ee
on ${\Bbb C}^4\otimes{\Bbb C}^4 \otimes{\Bbb C}^4 $, the subindices of
the $R$-matrix referring to the spaces on which $\Rc$ acts nontrivially;
ii) has a shift point, meaning that $\Rc(0)=\Bbb I$; and iii) is related
to the matrix representation $H$ through the relation
\begin{equation}
H= \sum_{i} {\Bbb I} \otimes {\Bbb I}\otimes \dots \otimes
\underbrace{\left.{\partial_u R} \right |_{u=0}}_{i,i+1}\otimes {\Bbb I}
\otimes \dots \otimes {\Bbb I}
\quad ,
\end{equation}
implying that the first order of $\Rc$ in the spectral parameter $u$ is
the representation of the (two-site) Hamiltonian.

In a recent paper \cite{DOMO}, it was shown that when the matrix
$\Rc$ is a finite polynomial of degree $p$ in $u$, $\Rc(u)={\Bbb I}+...
+ \frac{u^p}{p!} \Rc^{(p)}$, in order for the YBE to be satisfied the
coefficients in the polynomial expansion must themselves obey a certain
number of equations. In particular the highest degree coefficient order
$R^{(p)}$ must satisfy
\bq
\Rc_{23}^{(p)} \, \Rc_{12}^{(p)} \,\Rc_{23}^{(p)} &=& \Rc_{12}^{(p)}\,
\Rc_{23}^{(p)}\, \Rc_{12}^{(p)} \quad . \label{bgeq}\\
\left(\Rc_{}^{(p)} \right )^2 & \propto & {\Bbb I} \quad , \label{r2}
\eq
which are realized to be the braid group relations specialized to the
symmetric case $\Rc_{ij}=\Rc_{ji}$. This is also known as the symmetric
group. 

The simplest non-trivial case of a polynomial $R$-matrix is obtained
for $p=1$. In this case $\Rc(u)={\Bbb I}+u \Rc^{(1)}$, and, due to
(\ref{bgeq}), $\Rc^{(1)}$ is nothing but a braid operator. The
solutions of the braid group equation have been exhaustively investigated
for a two-dimensional local space (see for instance \cite{HIET}).
On the contrary, to our knowledge, little is known for the
4-dimensional case. Therefore in the next section we look for
$16 \times 16$ matrix solutions of eqns. (\ref{bgeq})-(\ref{r2})
parametrized in two (a priori) different forms: i) the generalized
permutators (to be defined in the following), ii) the extended Hubbard
matrices. Our results show that the latter class of solutions is
actually a subclass of the former, so that every integrable extended
Hubbard model deriving from a first order polynomial $R$-matrix is a 
generalized permutator.
 
It is well known that, among the solutions of the symmetric group
relations (\ref{bgeq})-(\ref{r2}), one always finds the permutation
operator $P$, defined by $P  (e_\alpha\otimes e_\beta) = e_\beta
\otimes e_\alpha$. In fact, this is true whatever the dimension of the
representation is, and also holds for any {\it graded} permutation
operator $P_g$, where $P_g (e_\alpha\otimes e_\beta)=
(-)^{\epsilon(\alpha) \epsilon(\beta)} e_\beta\otimes e_\alpha$.
Therefore, both $P$ and $P_g$ give rise to integrable models. In
particular, the model corresponding to the $R$-matrix having $P_g$ as
first degree coefficient is the $EKS$ Hamiltonian \cite{EKS}. One may
wonder whether there are other solutions of (\ref{bgeq}), (\ref{r2})
generalizing the structure of $P$, $P_g$.
Here we propose the {\it generalized permutator} ones, given by
operators $\Pi$ which either transform one product of basis vectors into
the reversed product, or leave it unchanged. In the matrix form this
means that there is precisely one non-zero entry in each column and row
of $\Pi$. Moreover, due to (\ref{r2}), the non-vanishing entries of
$\Pi$ must be equal to $+1$ or $-1$, up to an overall multiplicative
constant. Explicitly, $\forall \alpha,\beta =1, \ldots 4 $,
\[
\Pi (e_\alpha\otimes e_\beta) =
\epsilon_{\alpha\beta} \, \underbrace{s^{d}_{\alpha\beta} \,
(e_{\alpha} \otimes e_{\beta})}_{\mbox{\tiny diagonal terms}} \, + \,
(1-\epsilon_{\alpha\beta})  \underbrace{ s^{o}_{\alpha\beta} \,
(e_{\beta} \otimes e_{\alpha})}_{\mbox{\tiny off-diagonal terms}}
\nonumber\\ \,.
\]
Here $\epsilon_{\alpha\beta}$ is a discrete (0 or 1) valued function
satisfying  $\epsilon_{\alpha\beta}=\epsilon_{\beta\alpha}$ and
$\epsilon_{\alpha\alpha}=1$, which selects the diagonal/off-diagonal
terms. Moreover, $s^{d}_{\alpha\beta}=\pm 1$ accounts for additional
signs of the diagonal entries, while $s^{o}_{\alpha\beta}=\pm 1$ stands
for the signs of off-diagonal terms; we shall impose
$\,s^{o}_{\alpha\beta}=s^{o}_{\beta\alpha}$ in order to make $\Pi$ a
symmetric matrix.

The generalized permutator $\Pi$ has the form given by matrix
(\ref{gp}), where $\sigma^{d}_{\alpha\beta}=\epsilon^{}_{\alpha\beta}
s^{d}_{\alpha\beta} $ and $ \sigma^{o}_{\alpha\beta}=
(1-\epsilon^{}_{\alpha\beta}) s^{o}_{\alpha\beta}$, so that
$\sigma^{d}_{\alpha\beta} \cdot \sigma^{o}_{\alpha\beta}=0 \,\,
\forall \alpha,\beta$ ({\it i.e.} actually there is only one
non-vanishing element at each row/column, and its square equals +1).
\end{multicols}
\be
\Pi\,=\,{\scriptsize{\pmatrix{ 
\:\sigma^{d}_{11} & 0 & 0 & 0 &|&
0 & 0 & 0 & 0 &|&
0 & 0 & 0 & 0 &|&
0 & 0 & 0 & 0 &\cr
\:0 & \sigma^{d}_{12} & 0 & 0 &|& 
\sigma^{o}_{12} & 0 & 0 & 0 &|&
0 & 0 & 0 & 0 &|&
0 & 0 & 0 & 0 &\cr
\:0 & 0 & \sigma^{d}_{13} & 0 &|&
0 & 0 & 0 & 0 &|&
\sigma^{o}_{13} & 0 & 0 & 0 &|&
0 & 0 & 0 & 0 &\cr
\:0 & 0 & 0 & \sigma^{d}_{14} &|&
0 & 0 & 0 & 0 &|&
0 & 0 & 0 & 0 &|&
\sigma^{o}_{14} & 0 & 0 & 0 \cr 
- & - & - & - &|&
- & - & - & - &|&
- & - & - & - &|&
- & - & - & - &\cr 
0 & \sigma^{o}_{12} & 0 & 0 &|& 
\sigma^{d}_{21} & 0 & 0 & 0 &|&
0 & 0 & 0 & 0 &|&
0 & 0 & 0 & 0 &\cr
0 & 0 & 0 & 0 &|&
0 & \sigma^{d}_{22} & 0 & 0 &|&
0 & 0 & 0 & 0 &|&
0 & 0 & 0 & 0 &\cr
0 & 0 & 0 & 0 &|&
0 & 0 & \sigma^{d}_{23} & 0 &|&
0 & \sigma^{o}_{23} & 0 & 0 &|&
0 & 0 & 0 & 0 &\cr
0 & 0 & 0 & 0 &|&
0 & 0 & 0 & \sigma^{d}_{24} &|&
0 & 0 & 0 & 0 &|&
0 & \sigma^{o}_{24} & 0 & 0 \cr 
- & - & - & - &|&
- & - & - & - &|&
- & - & - & - &|&
- & - & - & - &\cr 
0 & 0 & \sigma^{o}_{13} & 0 &|& 
0 & 0 & 0 & 0 &|&
\sigma^{d}_{31} & 0 & 0 & 0 &|&
0 & 0 & 0 & 0 &\cr
0 & 0 & 0 & 0 &|&
0 & 0 & \sigma^{o}_{23} & 0 &|&
0 & \sigma^{d}_{32} & 0 & 0 &|&
0 & 0 & 0 & 0 &\cr
0 & 0 & 0 & 0 &|&
0 & 0 & 0 & 0 &|&
0 & 0 & \sigma^{d}_{33} & 0 &|&
0 & 0 & 0 & 0 &\cr
0 & 0 & 0 & 0 &|&
0 & 0 & 0 & 0 &|&
0 & 0 & 0 & \sigma^{d}_{34} &|&
0 & 0 & \sigma^{o}_{34} & 0 \cr 
- & - & - & - &|&
- & - & - & - &|&
- & - & - & - &|&
- & - & - & - &\cr
0 & 0 & 0 & \sigma^{o}_{14} &|& 
0 & 0 & 0 & 0 &|&
0 & 0 & 0 & 0 &|&
\sigma^{d}_{41} & 0 & 0 & 0 &\cr
0 & 0 & 0 & 0 &|&
0 & 0 & 0 & \sigma^{o}_{24} &|&
0 & 0 & 0 & 0 &|&
0 & \sigma^{d}_{42} & 0 & 0 &\cr
0 & 0 & 0 & 0 &|&
0 & 0 & 0 & 0 &|&
0 & 0 & 0 & \sigma^{o}_{34} &|&
0 & 0 & \sigma^{d}_{43} & 0 &\cr
0 & 0 & 0 & 0 &|&
0 & 0 & 0 & 0 &|&
0 & 0 & 0 & 0 &|&
0 & 0 & 0 & \sigma^{d}_{44} \cr 
}}} \label{gp}
\ee
\begin{multicols}{2}
Let us denote by ${\mathcal{N}}$ the set of index values, {\it i.e.}
${\mathcal{N}}=\{1,2,3,4\}$; the function $\epsilon_{\alpha\beta}$ on
${\mathcal{N}} \times {\mathcal{N}}$ characterizes the structure of any
generalized permutator. Let us denote by $\mathcal{D}$ the subset of
$\mathcal{N} \times \mathcal{N}$ of the couples $(\alpha,\beta)$ for
which $\epsilon_{\alpha\beta}=1$: they determine the positions
$4\,( \alpha-1)+\beta$ of the non-vanishing diagonal entries; since
$\epsilon_{\alpha\beta}=\epsilon_{\beta\alpha}$, whenever
$\sigma^{d}_{\alpha\beta} \neq 0$ we also have that
$\sigma^{d}_{\beta\alpha} \neq 0$; in other words, the subset
$\mathcal{D}$ can always be written in the form ${\mathcal{D}}=
\bigcup_{i} \, {\mathcal{E}}_{i} \times {\mathcal{E}}_{i}$ 
where the ${\mathcal{E}}_i$'s are disjoint subsets of $\mathcal{N}$.
\\Inserting a generalized permutator $\Pi$ into the braid group
equation (\ref{bgeq}) it can be easily found that $\Pi$ is a solution
if and only if $\sigma^{d}_{\alpha\beta}=S_{i} \quad \forall \,
(\alpha,\beta) \in {\mathcal{E}}_{i} \times {\mathcal{E}}_{i}$, where
$S_{i}=\pm 1$ are signs. The values of the $S_i$'s can be chosen
independently, and the remaining off-diagonal non-vanishing entries are
also free. In the following we shall refer to these solutions as the
{\it symmetric group generalized permutators} and denote them by $\Pi_s$.
\\As the above definition of $\Pi_{s}$ is independent of the dimension
of the representation, we conjecture that they are solutions of the
symmetric group relations in any dimension. However, their actual number
depends on the dimension of $V_j$. In the present case this is four,
and the number of different $\Pi_s$'s is 1440.

Let us now consider explicitly the possibility that the representation
of the two-site extended Hubbard Hamiltonian (up to an additive
constant $c$) is itself a solution of the symmetric group relations 
(\ref{bgeq})-(\ref{r2}). It turns out that actually there are 96
different possible
choices of values of the physical parameters in (\ref{ham}) satisfying
(\ref{bgeq})-(\ref{r2}). They can be cast into six groups as
(\ref{hsol}) shows.
\end{multicols}
\begin{equation}
\scriptsize{\matrix{
  &  &	H_1 (s_1,\dots, s_5) & | & H_2 (s_1,\dots,s_5) &|&
        H_3 (s_1,s_2,s_3)&| & H_4 (s_1,s_2,s_3) &|&
        H_5 (s_1,s_2,s_3)&| & H_6 (s_1,s_2,s_3) &|& \cr
  &||& ------& | &------& | & ------ & | & ------& | & ------ & | & ------
& |\cr
t &||&	1      &|&  1  &| &  1 &|      & 1 &|& 1  &|& 0  &|   \cr
X &||&	1      &|&  1  &| &  1 &|      & 1 &|& 1  &|& 0  &|  \cr
\tilde{X} &||&	1+s_2 &|& 1+s_2 & |& 1+s_2 &|&
1+s_2 &|& 1  &|& 1  &|\cr
U &||& 2 s_1&| & 2 s_1 &|& 4 s_1 &|& 4 s_1 &|& 2 s_1  &|& -2 s_1 &|  \cr
V &||& s_1 &|& s_1+s_4 &|& s_1   &|& s_1+s_3 &|& s_1+s_3  &|& 0 &|\cr
W &||& s_4 &|& 0 &|& s_3 &|& 0 &|& 0  &|& 0 &|\cr
Y &||& s_3 &|& s_3  &|  & 0 &|& 0 &|& s_2  &|& s_2 &|\cr
P &||& s_4-s_1 &|&  -s_1-2 s_4 &|& s_3-2 s_1 &|& -2 (s_1+s_3) 
&|& -(s_1+s_3) &|&  0  &| \cr
Q &||&  -2 s_4+s_1+s_5  &|& 4s_4+s_1+s_5 &|& 4 s_1-2 s_3 &|& 4 (s_1+s_3)
 &|&  s_1+s_3   &|&  s_1+s_3  &| \cr
\mu &||& -2 s_1&| & -2 s_1&| & -2 s_1&| & -2 s_1 &|& -2 s_1  &|& 0 &|
\cr
c &||& s_1&| & s_1&| & s_1&| & s_1 &|& s_1  &|& s_1 &| \cr
}} \label{hsol}
\end{equation}
\begin{multicols}{2}
\noindent Here $s_i=\pm 1$, $i=1,\dots, 5$ are arbitrary signs; the
first two groups consist of 32 different solutions each, while
any of the other four groups is made of 8 different cases.
\\One can notice that $t=X$ is a common feature exhibited by all the
solutions, implying that the number of doubly occupied sites is a 
conserved quantity for those ${\cal H}$ that are derivable from
first-degree polynomial $R$-matrices. This feature is important in that
it means that in these cases ${\cal H}$ can be diagonalized within a
sector with a given number of up and down electrons and doubly
occupied sites. In practice, the solvability of the model in one
dimension is not affected by having values of $U$ and $\mu$ in
${\cal H}$ other than those reported in (\ref{hsol}) (see also
\cite{EKS}). This is true also for the term proportional to the
identity: indeed the value $c=s_1$ exhibited by all the solutions just
stems from requiring that the $R$-matrix is a first degree polynomial.
\\\indent Remarkably, all solutions (\ref{hsol}) do belong to the class
of symmetric group generalized permutators $\Pi_s$. In fact, it is
easily checked that they are {\it all} those $\Pi_s$ which preserve
the spin ${\bf S}$ and the number of electrons ${\cal N}_e$, the latter
conditions reducing the 1440 integrable models identified by the
$\Pi_s$'s to precisely the 96 cases reported in (\ref{hsol}). In
other words, the integrable extended Hubbard Hamiltonians having a
first degree $R$-matrix are in one-to-one correspondence with the
symmetric group generalized permutators invariant under
$su (2)\oplus u(1)$.
\\One may wonder whether the 96 solutions (\ref{hsol}) have non-similar
$R$-matrices, {\it i.e.} whether they correspond to {\it independently}
integrable models. In the Yang-Baxter formalism, two matrices $\Rc$ and
$\Rc'$ are said to be similar if they are connected by a similarity
transformation of the form $\tau=A \, {\otimes}^s A$,
$A$ being defined on the site space and ${\otimes}^s$ being the graded
product\cite{BGLZ}; when this is the case, it is proven in general that,
for $\Rc$ satisfying YBE, $\Rc'$ fulfills the YBE as well \cite{KUSK},
the integrability of the latter being automatically deduced 
from the integrability of the former.
To answer the above question, we first observe that the traces of the various 
solutions can be different, assuming even values from $-10$ to $10$.
Indeed, for the first four groups, the trace is given by the value of
$Q$, whereas for the last two groups it is given by $9 s_3+s_1$ and
$9 s_1+s_3$ respectively. It is then obvious that solutions with
different trace values can not be similar, so that the above question
needs to be answered only within a class of given trace value. Let us
focus for instance on the models with vanishing trace. They are 12
and all have $Q=P=0 \,$; 8 of them belong to the first
group and are characterized by the Hamiltonian
$H_1 (s_1,s_2,s_3,s_1,s_1)$, whereas the other 4 are subcases of the
fourth group with $H_4 (s_1,s_2,-s_1)$. We have verified that no
pairs of their $R$-matrices are similar, so that the 12 models
with vanishing trace are independently integrable models.
\\However, there are other kinds of unitary transformations mapping some of
the solutions (\ref{hsol}) into other ones. This feature stems from the
invariance properties of the Hamiltonian ${\cal H}$, which is mapped into
itself by means of the canonical transformations $c_{i, \sigma} \rightarrow
(-)^{i} c_{i,\sigma}$, and $c_{i, \sigma} \rightarrow [1-(1-(-)^{i})
n_{i, -\sigma}]
c_{i, \sigma}$, yielding 
${\cal H} \,(t,X,\tilde X,U,V,W,Y,P,Q,\mu)\stackrel{U^{(a)}}
{\longrightarrow}
-{\cal H}\, (t,X,\tilde X,-U,-V,$ $-W,-Y,-P,-Q,-\mu)$ and 
${\cal H} \,(t,t,$ $\tilde X,U,V,W,Y,P,Q,\mu)$ $\stackrel{U^{(b)}}
{\longrightarrow}
{\cal H}\, (t,t,2 t-\tilde X,U,V,$ $W,-Y,P,Q,\mu)
$ respectively.

Within the various groups one can recognize some known integrable
models, all having $P=0=Q$, $t\neq 0$. 
The EKS hamiltonian \cite{EKS} is $H_1 (-1,-1,-1,-1,-1)$ which in fact
corresponds to the standard graded permutator $P_g$,
whereas $H_1(1,-1,1,1,1)$ is the EKS model with opposite grading.
The AAS hamiltonian \cite{ALAR} is nothing but $H_4 (-1,-1,1)$,
whereas in the fifth group $H_5 (s_1, s_2, s_1)$ are particular cases
of the supersymmetric $U$ model \cite{BGLZ}, whose most general form is
however associated with a second degree polynomial $R$-matrix
\cite{DOMO}. It is worth stressing that even for the more standard case 
$P=0=Q$, $t\neq 0$, the integrable models presented here, amounting
to 16, are a far wider class than those which can be found in the
literature.

Once for a given model the $R$-matrix is known, its spectrum is obtained
within QISM by diagonalizing the corresponding Transfer matrix. In
practice here the exact solutions, the details of which are deferred to a
future publication, is achieved by closely following the Bethe Ansatz
analysis for Hamiltonian acting as permutators, first considered by
Sutherland in \cite{SUT} (see also \cite{EKS} for its application to the
EKS model). Let us just mention one physically interesting feature of
the phase diagram which emerges in all the solutions characterized by
$V={1\over 2}(W\pm Y)$, and survives also in dimension greater than 1,
as shown in \cite{dBKS}: it is the appearence, below a critical value
of $U$, of a superconducting phase characterized by
$\eta$-pairs\cite{YAN,MOCA}.
\\Finally, it is worth noticing that the solutions to equations
(\ref{bgeq})--(\ref{r2}) presented here are expected to be relevant
also in the search for completely integrable models with higher degree
polynomial $R$-matrices. Indeed they provide a class of possible
coefficients $\Rc^{(p)}$ for the highest degree term, in the form of
symmetric group generalized permutators. Work is in progress along these
lines.

\end{multicols}

\end{document}